\documentclass[aps, floats, twocolumn, superscriptaddress]{revtex4}
\usepackage[final]{graphicx}
\DeclareGraphicsExtensions{.eps,.ps,.pdf}
\usepackage{amsmath,amssymb}
\usepackage{bm}

\newcommand\vex[1]{\mathbf{#1}}
\newcommand\dbar[2]{\frac{\d^{#1}#2}{(2\pi)^{#1}}\,}

\def\d{\mathrm{d}}
\def\id{\openone}
\def\tr{\mathrm{tr}}

\begin{document}

\title{Excitons in QED$_3$ and spin response in a phase-fluctuating $d$-wave superconductor}
\author{Babak H. Seradjeh}
\email{babak@physics.ubc.ca}
\affiliation{Department of Physics and Astronomy, University of British Columbia,
Vancouver, BC, Canada V6T 1Z1}
\author{Igor F. Herbut}
\affiliation{Department of Physics, Simon Fraser University,
Burnaby, BC, Canada V5A 1S6 }

\begin{abstract}
We study the particle-hole exciton mode in the QED$_3$ theory of a phase-fluctuating $d$-wave superconductor in ladder approximation. We derive a Schr\"odinger-like equation for the exciton bound state and determine the conditions for its existence. We find the dispersion of this mode below the particle-hole continuum and compare our results with the resonance observed in neutron scattering measurements in cuprates.
\end{abstract}


\typeout{polish abstract}
\maketitle

\section{Introduction}
Strong electronic correlations in high-temperature superconductors
are believed to underlie their rich phenomenology. The
superconducting gap in these materials exhibits $d$-wave symmetry
with four nodes. Since the gap vanishes and changes sign at these
nodes, the low-energy fermionic excitations have linear, Dirac-like,
dispersion. An important probe of the correlations of these nodal
quasiparticles is provided by inelastic neutron scattering
measurements, which give us information about the momentum and
energy dependence of the correlations of electronic spin degrees of
freedom through their coupling to scattering neutrons. Over the past
decade a consistent experimental picture of these correlations has
been emerging in different families of cuprates, so far the
exclusive hosts of high-temperature superconductivity. These include
the
double-layer~\cite{FonBouSid99a,HeSidBou01a,FonKeiAnd95a,HayMooDai04a,StoBuyLia04a,StoBuyCow05a,HinBouPai06a}
Bi- and Y- and the
single-layer~\cite{YamWakShi95a,YamLeeKur98a,ChrMcMRon04b,TraWooPer04b,HeBouSid02a,BouKeiPai05a}
La- and Tl-based families. The salient features of this picture are
the following: (1) A resonance peak at the antiferromagnetic
ordering wave vector, $(\pi,\pi)$, at a resonance energy
$\omega_{\rm res}$ ($\approx41$~meV at optimal doping); (2) A
two-dimensional~\cite{HinPaiBou04a} incommensurate structure
\emph{below} the resonance energy with maxima located at
$(\pi\pm\delta,\pi)$ and with a possible weak inward dispersion
toward $(\pi,\pi)$ as $\omega\to\omega_{\rm res}$; (3) An
incommensurate structure \emph{above} the resonance with outward
dispersion away from $(\pi,\pi)$ for higher energies. The above
dispersions have been taken to suggest that the commensurate and
incommensurate peaks may have a common origin.

In addition, the response is usually not discernible from the
background below a certain energy,  referred to as the ``spin gap.''
However, for the sake of clarity, we reserve the label ``spin gap''
strictly for the minimum value of particle-hole continuum,
$\omega_{\rm sg}$, and call the former ``spin response threshold.''
Thus, whether the observed signals at $(\pi,\pi)$ and $\omega_{\rm
res}$, and/or incommensurate peaks are true resonances is closely
linked to the question of whether they occur below the corresponding
spin gap. Accordingly, one could think of two possible theoretical
scenarios.

In the first scenario, there is no true resonance in the system,
that is, no particle-hole bound state, and the peaks are merely the
maxima of the spin response.  The nodal quasiparticles will have a
finite spin gap everywhere except at the nodes. At $(\pi,\pi)$, the
peak would be interpreted as the overlap of the four incommensurate
responses, one for each node, at the center. The spin gap and the
spin response threshold are the same in this scenario.  Usually this
should also mean that the central peak falls off slowly as
$1/\omega$, rather than being sharp. However, as noted above, the
central peak observed in experiment seems to be very sharp, in some
cases limited only by the energy resolution of the measurement.

In the second scenario, there would be a true resonance below the
spin gap given by a $\delta$-function in the spin response, due to
the formation of particle-hole bound states---the so-called spin
excitons. In this scenario, the spin gap is distinct from the spin
response threshold. It would still be important to determine the
continuum response so we could decide what effects derive from which
source. However, it would also be very important to determine the
existence (or lack of) and the properties of such excitons in a
candidate theory of cuprates.

In this paper, we study the existence of spin  excitons within the
QED$_3$ effective theory of underdoped
cuprates~\cite{FraTes01a,Her02a,Her02b,FraTesVaf02a}. The QED$_3$
theory describes a $d$-wave superconductor in which the phase of the
superconducting order parameter is fluctuating. These phase
fluctuations are encoded in a pair of $U(1)$ gauge fields, the Berry
gauge field and the Doppler gauge field, that couple to
quasiparticle's spin and charge degrees of freedom, respectively.
Both gauge fields are massive in the superconducting phase, which,
accordingly, exhibits sharp nodal excitations. As the system is
underdoped, however, the phase-fluctuations grow stronger and
eventually destroy the phase coherence of the superconducting order
parameter. The resulting state may be shown to be insulating
~\cite{Her05a}. Whereas outside the superconductor the Doppler gauge
field remains massive, the Berry gauge field becomes massless. This
allows the general chiral symmetry breaking instability  in the
QED$_3$ ~\cite{AppNasWij88a,Mar96a} to become operative, which in
the present context implies the antiferromagnetic ordering in the
system~\cite{Her02a,SerHer02a}.

Herbut and Lee~\cite{HerLee03a} have previously calculated the
QED$_3$ spin response in the superconducting phase in a low-energy
approximation and found no exciton resonance. In their approach the
Berry gauge-field propagator in the superconducting phase was
approximated by a constant, equal to its infrared mass. This result
falls therefore into the first scenario discussed above. This ``constant-mass'' approximation is valid below a momentum cut-off that is proportional to the gauge-field mass. So, it becomes less reliable in the underdoped region where the mass decreases and eventually vanishes at the underdoped transition. In the
present paper we go beyond the constant-mass approximation by
including the momentum-dependence of the inverse propagator of the
Berry gauge field, which becomes linear at high momenta. We do so by considering the scalar vertex for the spin response in a ladder
approximation and by reducing the problem to an approximate
Schr\"odinger equation that describes the formation of particle-hole
bound states in the $d$-wave superconductor. We find that,
for a strong enough coupling between the nodal quasiparticles
mediated by the Berry gauge field such bound states exist. We derive
their dispersion and compare our results with the experimental
picture outlined above. However, due to the inherent gauge dependence of our ladder approximation we cannot definitely determine whether, for the physical values of parameters, QED$_3$ theory of cuprates is in this strong-coupling regime. 

We note that the resonance is absent in La-cuprates. This could be
due to different physics from competing stripe order that is known
to exist in this family and results in a smaller critical
temperature, $T_c$. Alternatively, within our theory, one might
expect the resonance to be harder to discern from the incommensurate
structure for a smaller $T_c$. See Sec.~\ref{sec:conc} for a
discussion.

The paper is organized as follows: In Sec.~\ref{sec:ladder}  we
formulate the ladder approximation for the scalar vertex and the
spin response. In Sec.~\ref{sec:exciton} we derive the Schr\"odinger
equation for excitons, find the resulting expression for resonant
spin response, and determine the conditions for their existence. In
Sec.~\ref{sec:exp} we discuss the dispersion of the excitons and
compare with experiments and existing literature. We summarize our results in Sec.~\ref{sec:conc} where we also comment on the issue of gauge dependence mentioned above. The details of some of our calculations are
given in two appendices.

\section{Ladder approximation}\label{sec:ladder}
The spin part of the effective action for the nodal quasiparticles
in the fluctuating d-wave superconductor~\cite{Her05a} is given
by
\begin{eqnarray}
S = \int\d^3 r\sum_{i=1}^{N} \bar\psi_i \gamma_\mu\left(v_{i,\mu}
\partial_\mu-iga_\mu\right) \psi_i \nonumber \\
\label{eq:Action} 
+ \frac12 \int\dbar3p a_{\mu}(-p)D_{\mu\nu}^{-1}(p)a_{\mu}(p),
\end{eqnarray}
where the number of flavours, $N=2$, is the number of pairs of nodes.
The bare value of the coupling $g$ is unity. The Dirac spectrum is
assumed for quasiparticles near the nodes with anisotropic
velocities $v_{1,\mu}=(1,v_F,v_\Delta)$ and
$v_{2,\mu}=(1,v_\Delta,v_F)$. Here $v_F/v_\Delta\sim10$ is the ratio
of the Fermi velocity to the gap gradient at the node. We will set
$v_F=v_\Delta$ throughout our calculations and restore their values
by rescaling the corresponding momenta at the end. Only the
transverse components of the gauge field enter the action in
Eq.~(\ref{eq:Action}):  The gauge-field propagator has the form
$D_{\mu\nu}(p)=\big(\delta_{\mu\nu}-(1-\xi)\hat p_\mu\hat
p_\nu\big)D(p)$, where~\cite{HerLee03a}
\begin{eqnarray}\label{eq:Dp}
D(p) &=& \frac{\pi}{4|p|}F(\frac{m}{|p|}), \\
F(z) &=& (4z^2+1)\tan^{-1}\frac1{2z}-2z,
\end{eqnarray}
and $\xi$ signifies a (nonlocal) gauge-fixing term~\cite{foot1}.  The gauge
field has a mass $D^{-1}(0)=12m/\pi$. At high energies,
$D^{-1}(p)\sim|p|$, which is an exact result~\cite{HerTes96a}. Most
of what we will say about excitons in QED$_3$ does not depend on the
exact form of the interpolation between the infrared mass and the
ultraviolet linear dependence of $D^{-1}(p)$, but for the sake of
definiteness we will present our results for this particular form of
the propagator, which is found from the dual theory of
Refs.~\onlinecite{LeeHer03a} and~\onlinecite{Her05a}.

The spin operator, $S_z$, is related to the Dirac fields
through~\cite{Her02a},
$$
\bar\psi_i(r)\psi_i(r) = 4\cos(2\vex K_i\cdot\vex r)S_z(r),
$$
where $\vex K_{1,2}=(\pm k_F,k_F)$ denote the positions of two of
the the nodes in the Brillouin zone. We assume spin-rotational symmetry. Thus,
$\chi_{\mu\nu}(\vex k, \omega) = \left< S_\mu(-\vex k,
-\omega)S_\nu(\vex k, \omega) \right>  = \delta_{\mu\nu}\chi(\vex k,
\omega), \nonumber$ where $\chi(\vex k, \omega) = \left< S_z(-\vex
k, -\omega)S_z(\vex k, \omega) \right>.$ Writing $\vex k=2\vex
K_i+\vex p$, we find
\begin{equation}
\chi(p) = \frac1{16}\int\d^3r e^{-ip\cdot r}\sum_{ij}\left< \bar\psi_i(r)\psi_i(r)\bar\psi_j(0)\psi_j(0) \right>.
\end{equation}

\begin{figure}[t]
\includegraphics[scale=0.4]{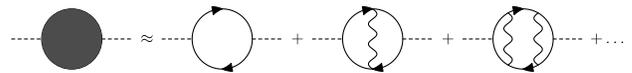}
\caption{\label{fig:ladderX}Ladder approximation for the spin response}
\end{figure}

In order to calculate the spin response in  the QED$_3$ action we
will adopt a  ladder approximation for the four-point correlator
$\chi(p)$, shown diagrammatically in FIG.~\ref{fig:ladderX}. The
first diagram represents the ``bare'' spin response in the absence
of gauge-field interactions. We denote it by $\chi_{0}(p)$.

\begin{figure}[t]
\includegraphics[scale=0.4]{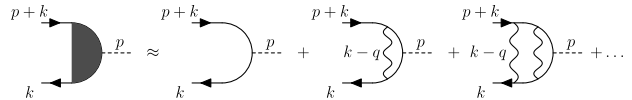}
\caption{\label{fig:ladderV}Ladder approximation for the scalar vertex, $\Gamma(k,p)$.}
\end{figure}

It is standard to reformulate the ladder approximation in terms of
gauge-field-induced corrections  to the scalar vertex for the
interaction between the external source (neutrons) and spinons.
These vertex corrections are shown in FIG.~\ref{fig:ladderV}.
Defining the scalar vertex, $\Gamma(k,p)$, as the amputated diagram
(without the external legs) on the left-hand side, we have
\begin{equation}\label{eq:Xladder}
\chi(p) = -\frac{N}{16} \int\dbar3k \tr\left[ G_0(k)\Gamma(k,p)G_0(k+p) \right],
\end{equation}
where $G_0(k) = {-i\gamma_\mu k_\mu}/{k^2}$ is the bare
four-component spinon  propagator in the superconducting state. From
FIG.~\ref{fig:ladderV} we see that the scalar vertex satisfies the
following Bethe-Salpeter equation,
\begin{eqnarray}
\Gamma(k,p) = \id - g^2\int \dbar3q \gamma_\mu G_0(q)\Gamma(q,p)G_0(q+p) \nonumber \\ \label{eq:BS}
\times \gamma_\nu D_{\mu\nu}(k-q).
\end{eqnarray}

We may use the symmetries of the QED$_3$ action to expand the scalar
vertex in terms of form factors~\cite{Bur92a},
\begin{eqnarray}
\Gamma(k,p) &=& \id F_1(k,p) +  \sigma_{\mu\nu} k_\mu p_\nu F_2(k,p) \nonumber \\
&& +\gamma_\mu k_\mu F_3(k,p)  + \gamma_\mu p_\mu F_4(k,p).
\end{eqnarray}
Here, $\sigma_{\mu\nu}=\frac12[\gamma_\mu,\gamma_\nu]$. Then, Eq.~(\ref{eq:BS}) is reduced to
\begin{equation}\label{eq:formfK}
F_i(k,p) = \delta_{i1} + \int \dbar3q \sum_{j=1}^4 K_{ij}(k,q,p) F_j(q,p),
\end{equation}
with the matrix $K_{ij}$ having a block-diagonal form in terms of
two two-by-two matrices. For the reasons explained below the most
important element is
\begin{equation}
K_{11}(k,q,p) = (2+\xi)g^2\frac{q\cdot(q+p)}{q^2(q+p)^2}D(k-q).
\end{equation}
The block-diagonal form of $K_{ij}$ implies that the form factors
$F_3$ and $F_4$ completely decouple from $F_1$ and $F_2$.  Together
with Eq. (6), which implies that $F_3$ and $F_4$ do not contribute
to the function $\chi$, this allows us then to neglect them
altogether for the purpose of calculating the spin response.

Within the ladder approximation for the scalar vertex,  our
manipulations have so far been exact. At this point, we notice that
the dominant contribution to $\chi$ comes from
$F_1$, which  is of order $g^0$. Since $F_2 \sim
g^2$, we will then neglect $F_2$ to obtain the set of
equations~\cite{GusRee98a,GusHamRee01a},
\begin{eqnarray} \label{eq:XF1} 
\chi(p) &=& \frac{N}4\int\dbar3k \frac{k\cdot(k+p)}{k^2(k+p)^2}F_1(k,p), \\ 
F_1(k,p) &=& 1 + \lambda \int\dbar3q \frac{q\cdot(q+p)}{q^2(q+p)^2}D(k-q) F_1(q,p), \nonumber \\
\label{eq:F1}
\end{eqnarray}
for the spin response, with $\lambda=(2+\xi)g^2$ as the
(gauge-dependent) coupling strength. These equations provide the
basis for our further study of the exciton modes in QED$_3$. For a constant $D(k-q)$ they reproduce the results of Ref.~\onlinecite{HerLee03a}.

Note that Eqs.~(\ref{eq:F1}) and~(\ref{eq:XF1}) explicitly depend on
the choice of gauge.  This arises because the ladder diagrams in
FIG.~\ref{fig:ladderV} are not gauge-invariant if the bare fermion
propagator is used. The gauge invariance can be restored by choosing
the (non-local) gauge-fixing parameter $\xi$ so that the bare
fermion propagator satisfies the usual Ward-Takahashi identity with
the vertex. In principle this procedure will yield a
momentum-dependent function, $\xi(p)$~\cite{GusHamRee01a}. With
$m=0$, we find Nash's gauge $\xi=2/3$~\cite{Nas89a} whereas in the
opposite limit of constant-mass approximation one finds Feynman's
gauge, $\xi=1$. In the rest of the paper, we choose to work with a
momentum-independent gauge-fixing parameter $\xi$ for simplicity.

\section{Excitons in QED$_3$}\label{sec:exciton}
Our strategy to solve the vertex function is to obtain a (set of)
differential equation(s) from the integral equation~(\ref{eq:F1}).
This is similar to the classic derivation of exciton bound states in
metals and semiconductors~\cite{Mah67c,Mah67a}. We will show that
there is an approximate Schr\"odinger-like equation that captures
the bound-state content of the vertex function. Similar
Schr\"odinger equations have been derived before for the
non-relativstic limit of vertex functions in
QED~\cite{CasLep78a,AllBur96a}. Based on this Schr\"odinger
equation, we will determine the conditions for the existence of the
exciton bound states.

\subsection{Schr\"odinger equation}
Let us define a generalized response
\begin{equation}
\Phi(r,p) = \sigma_\mu\sigma_\nu\int\dbar3k \frac{k_\mu(k+p)_\nu}{k^2(k+p)^2}e^{ik\cdot r}F_1(k,p),
\end{equation}
where $\sigma_\mu=(\sigma_1,\sigma_2,\sigma_3)$ are Pauli matrices
(up to cyclic permutations).  Then Eqs.~(\ref{eq:XF1})
and~(\ref{eq:F1}) may be written as
\begin{eqnarray}
\chi(p) &=& \frac{N}4\tr\left[\Phi(0,p)\right] \equiv \frac{N}4\phi(0,p), \\
\label{eq:F1Phi}
F_1(q,p) &=& 1+\lambda\int\d^3r e^{-iq\cdot r}D(r)\phi(r,p).
\end{eqnarray}
Note that $r$ is a real-space variable and $D(r)$ indicates the real-space gauge-field propagator.

We see that
\begin{equation}\label{eq:UpSchr}
-\big[(ip+p\times{\bm{\sigma}})\cdot\partial+\partial\cdot\partial)\big]\Phi(r,p)
= \delta^{(3)}(r) + \lambda D(r)\tr\left[\Phi(r,p)\right].
\end{equation}
We may shift away the momentum $p$ (which enters the equation only
as a parameter) in the  derivative by completing the square. This is
achieved by a phase transformation of the form,
\begin{equation}\label{eq:trans}
\Phi \to \Phi' = \exp\left[\frac12(ip+p\times{\bm{\sigma}})\cdot r \right]\Phi.
\end{equation}
We note that $\phi'(0,p) = \phi(0,p)$. The transformed equation is, then,
\begin{equation}\label{eq:Up'Schr}
\big[-\partial\cdot\partial+(p/2)^2\big]\Phi'-\lambda
D(r)e^{\frac12{\bm{\sigma}}\cdot L}\tr\big[
e^{-\frac12{\bm{\sigma}}\cdot L}\Phi' \big] = \delta^{(3)}(r),
\end{equation}
Where we have denoted the orbital angular momentum of the
particle-hole  system by $L=r\times p$. This is an exact mapping of
the integral equation~(\ref{eq:F1}) to a set of coupled
Schr\"odinger-like differential equations.

Even though we have succeeded in deriving a set of differential
equations   for our vertex function, they are still very
complicated. However, for the purpose of studying the bound-state
content of the vertex function, we may simplify these equations
further down to a single Schr\"odinger-like equation. A numerical
solution seems to be the only useful alternative. As shown in
Appendix~\ref{app:details}, this approximate, decoupled equation is
given by
\begin{equation}\label{eq:apprSchr}
\big[(p/2)^2 - \partial\cdot\partial - V_{\rm eff}(r,p,m)\big]\phi'(r,p) = \delta^{(3)}(r),
\end{equation}
where, for $0<|p|<2m$,
\begin{equation}\label{eq:VeffIR}
V_{\rm eff} = \lambda\left(1-\frac{\lambda\sin^2\theta}{16}\right)\cosh^2|L/2|\:D(r).
\end{equation}
Here, $\theta=\cos^{-1}(\hat p\cdot\hat r)$. We expect
Eq.~(\ref{eq:apprSchr})  to provide a valid description of the bound
states for an adequately short-ranged interaction $D(r)$. This is
the case when $0<|p|<2m$. For $|p|>2m$, Eq.~(\ref{eq:VeffIR}) is valid only at short
distances. At large distances $r\gg (|p|-2m)^{-1}$, the potential becomes infinite for
$\theta>\sin^{-1}(2m/|p|)$:
\begin{equation}\label{eq:apprSchrL}
V_{\rm eff}\sim -\frac{\lambda^2}{p^2}\sinh^2|L|\:D^2(r)\to-\infty.
\end{equation}
We note that in the case of constant-mass approximation~\cite{HerLee03a} for
which  $D(r)\sim\delta^{(3)}(r)$, and also for $p=0$ in the general
case, Eq.~(\ref{eq:apprSchr}) follows exactly from
Eq.~(\ref{eq:Up'Schr}) with $V_{\rm eff}(r)=\lambda D(r)$.

\subsection{Resonant spin response}
We may now find $\chi(p)$ by solving the following three-dimensional
Schr\"odinger equation for  the eigenvalue $e_n(|p|,m)$ and the
normalized eigenfunction $\psi_n(r,|p|,m)$,
\begin{equation}\label{eq:excitonSchr}
\left[ -\partial\cdot\partial - V_{\rm eff} \right]\psi_n = e_n\psi_n.
\end{equation}
Then,
\begin{equation}
\chi(p) = \frac{N}4\phi'(0,p) = N\sum_n\frac{|\psi_n(0,|p|,m)|^2}{4e_n(|p|,m)+p^2}.
\end{equation}
By analytically continuing to real frequencies, $p_0\to
-i\omega+0^+$, and denoting $p_M=\sqrt{\vex p^2-\omega^2}$, we find that since $\Im e_n(p_M,m)=0$ the imaginary part of the spin
response observed in experiment is given by
\begin{eqnarray}\label{eq:excitonImX}
\Im\chi(p_M) = N\sum_n\pm|\psi_n(0,p_M,m)|^2 \delta\left(p_M^2+4e_n(p_M,m)\right).
\end{eqnarray}
where the sign $\pm=\mathrm{sgn}(\omega)\mathrm{sgn}\left(1+\partial e_n(p_M,m)/\partial p_M^2\right)$. If $e_n(p_M,m)$ were not real there would be no $\delta$-function, hence no resonance.

From Eq.~(\ref{eq:excitonImX}), we see that the necessary and
sufficient condition for the existence  of excitonic resonances at
$\vex p$ and $\omega<|\vex p|$ is that our Schr\"odinger
equation~(\ref{eq:excitonSchr}) admits bound state solutions with
real and negative eigenvalues, $e_b$, that solve the equation
\begin{equation}\label{eq:excitonBound}
e_b(p_c,m) = -\frac{p^2_c}4.
\end{equation}
We will now study the existence of such bound states.

\subsection{Existence of excitons}\label{sec:exist}
As shown in Appendix~\ref{app:rsD} the gauge-field propagator $D(r)$ scales as $1/r^2$. For $|p|>0$, we may rescale the space as $|p|r\equiv z$ to see that the energy spectrum satisfies the scaling relation
\begin{equation}
e_b(|p|,m)=p^2\varepsilon_b\left(2m/|p|\right).
\end{equation}
The scaling function $\varepsilon_b$ is the bound-state energy eigenvalue for the rescaled Schr\"odinger equation,
\begin{equation}
\left[ -\nabla_z^2 - \tilde V_{\rm eff}(z,\mu) \right]\psi_b(z,\mu) = \varepsilon_b(\mu)\psi_b(z,\mu),
\end{equation}
where $\mu=2m/|p|$ and the rescaled potential is
\begin{equation}
\tilde V_{\rm eff}(z,\mu) = \frac1{p^2}V_{\rm eff}\left(\frac z{|p|},1,\frac{\mu|p|}2\right).
\end{equation}
The potential $\tilde V_{\rm eff}$ has an inverse-square form for small
$z$.  The condition~(\ref{eq:excitonBound}) for the resonant
response is satisfied for $\varepsilon_b(\mu_c)=-1/4$; then,
$p_c=2m/\mu_c$.

The inverse-square potential is an instance of conformal anomaly,
i.e., the breakdown of scale symmetry in quantum mechanics. The
potential $-\lambda/16z^2$ has an infinite number of negative
energies for $\lambda>\lambda_0^*=4$ and is unbounded from
below~\cite{LanLif81a}. Thus, in this ``strong-coupling''
regime, the problem needs to be renormalized for it to be physically
meaningful. Although there are different ways of doing so, the
result is unique~\cite{CamEpeFan00a}. The inverse-square potential
has a single renormalized bound-state. The energy of the
renormalized bound-state cannot be determined within the theory; it
is an input of the theory either from experiment, or from the
physics at higher energies, beyond the domain of physical validity
of the potential. The continuum spectrum that is renormalized into a
single bound-state is called the ``conformal tower." The bound-state
wave function is given by
$$
\psi_0(z) = \sqrt{\frac{\kappa ^3}{2\pi}}\frac{K_0(\kappa |z|)}{\sqrt{\kappa |z|}},
$$
where $-\kappa^2$ is the renormalized bound-state energy, and $K_0$ is the modified Bessel function of the second kind.

For $\mu\geq 1$ the potential $\tilde V_{\rm eff}$ decays exponentially at
large distances.  However, the inverse-square form at small $z$ is
expected to be sufficient for the existence of a conformal tower in
the strong coupling $\lambda>\lambda^*$. By using $\psi_0$ as a trial wave function to calculate the energy one can check that this expectation is met for
$\lambda^*/\lambda^*_0\approx1.268$. The corresponding critical
charge is given by $g_c \approx 2.252/\sqrt{2+\xi}.$ In Nash's and
Feynman's gauge, for example, one finds $g_{c,N}=1.379$ and
$g_{c,F}=1.300$, respectively. The strong-coupling regime is found for $g>g_c$.

For $0\leq \mu<1$, the potential $\tilde V_{\rm eff}\sim -e^{2(\sin\theta-\mu)|z|}\to-\infty$ for $\theta>\sin^{-1}\mu$. So it is less clear whether the conformal tower would  exist in this limit. However, the same way as above and by choosing $\kappa>(1-\mu)/2$, we have checked that  the conformal tower still persists for the same strong-coupling regime. This result reflects the fact that the conformal tower for bound-states is essentially a short-distance phenomenon produced by the singular behaviour of the potential at the centre.

\section{Discussion and comparison with experiment}\label{sec:exp}
In this section we will assume that QED$_3$ theory is in the strong-coupling regime characterized in Sec.~\ref{sec:exist} and investigate the consequences of this assumption for the spin response observed in neutron scattering measurements.

\begin{figure}[t]
\includegraphics[scale=0.7]{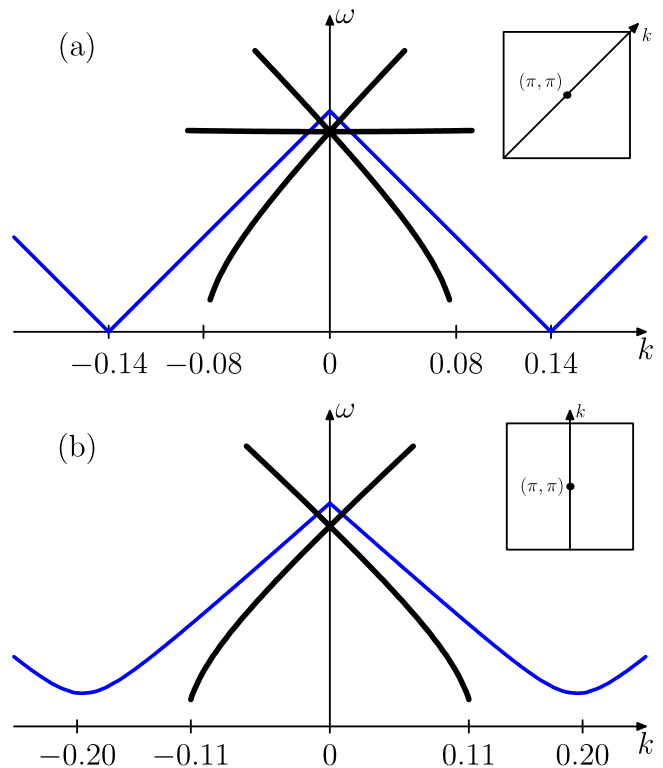}
\caption{\label{fig:disp}(Color online) The dispersion branches of the exciton in (a) the diagonal direction; and (b) the parallel direction, as shown in the insets.  The wave vectors are in reciprocal lattice units, and the origin is at $(\pi,\pi).$ The value of parameters used are $v_\Delta/v_F=0.15$, $k_F=0.4\pi$ and $p_c/v_F=\pi/16$. The continuum spin gap is shown by the thick blue line and vanishes at the nodes.}
\end{figure}

The dispersion of the exciton mode can be read off from Eq.~(\ref{eq:excitonImX}) and Eq.~(\ref{eq:excitonBound}) to be
\begin{equation}\label{eq:disp}
\omega_{\rm res}(\vex p) = \sqrt{\omega_{\rm sg}^2(\vex p)-p_c^2},
\end{equation}
where $\omega_{\rm sg}(\vex p)$ is the ``spin gap,'' given by the minimum of the particle-hole continuum. With the linear Dirac spectrum for spinons, it is
$$
\omega_{\rm sg}(\vex p) = v_F^2p_x^2+v_\Delta^2p_y^2,
$$
where we have now restored the Fermi and gap velocities. In this
formula, $p_x||v_F$ is the component  of the wave vector measured
from the given node in the nodal direction and $p_y||v_\Delta$ is
the one in the perpendicular direction. There is one dispersion
branch for each node. Due to the strong anisotropy
$v_\Delta/v_F\ll1$, the spin gap is much more sensitive to the
changes of $\vex p$ in the nodal direction. This causes the
dispersion~(\ref{eq:disp}) to assume a characteristic shape shown in
FIG.~\ref{fig:disp}. This is in qualitative agreement with
experiment if we identify the exciton mode with the observed
resonance peak~\cite{ChrMcMRon04b}. We will now discuss some
features of the dispersion.

In the diagonal direction, FIG.~\ref{fig:disp}(a),  there are three
exciton branches. Each of the nodes on the same diagonal line
contribute one branch, and the other two nodes produce overlapping
branches.  They all cross at $(\pi,\pi)$ and hence a strong
resonance is expected here. The overlapping branch is rather flat.
Its curvature is determined by the ratio $v_\Delta/v_F$. It quickly
enters the continuum away from $(\pi,\pi)$ and is, then, presumably
damped. The other two branches fall off rather steeply (due to the
square root in $\omega_{\rm res}$) and terminate a distance
$p_c/v_F$ away from the corresponding nodes.

In the parallel direction, FIG.~\ref{fig:disp}(b), there are only
two sets of overlapping branches,  each from a pair of nodes that
map to each other upon reflection about the parallel line. Here,
too, we observe a crossing at $(\pi,\pi)$ and, therefore, an
enhanced response. The absence of the flat branch means that the
momentum width of the resonance at $(\pi,\pi)$ is only bound by the
momentum resolution of the experiment in this direction. Again, the
modes disperse down to zero energy at an incommensurate position
$(\pi\pm\delta_{\rm inc},\pi)$ where
\begin{equation}\label{eq:dinc}
\delta_{\rm inc}=\frac{2(\pi-2k_F)-\sqrt2p_c/v_F}{1+v_\Delta^2/v_F^2}\lesssim2(\pi-2k_F)-\sqrt2p_c/v_F,
\end{equation}
and the approximation is made for $v_\Delta/v_F\ll1$.

In our Schr\"odinger equation approximation, the strength of the
peak is  given by $|\psi_n(0,p_c)|^2$  at the resonance ($p_M=p_c$)
and is the same for different momenta. That is, the whole branch
below the continuum spin gap has the same intensity. This does not
agree with experiments in which the resonance peak appears to go
away for low energies.

We also briefly note the effects of doping, $x$, on the dispersion
of  the exciton. In our formulation, doping enters through the
dependence of the gauge-field mass, $m(x)$, the position of the
nodes, $k_F(x)$, and the spin gap, $\omega_{\rm sg}(\vex p, x)$. The
gauge-field mass scales with $T_c$~\cite{HerLee03a,Her05a}. Thus, it
decreases with underdoping and vanishes at the underdoped
superconducting transition. As we approach half-filling, the nodal
points move towards $(\pi/2,\pi/2)$ and $k_F$ increases upon
underdoping. This causes the spin gap to decrease~\cite{Ser06a}. 
From our scaling relations in Sec.~\ref{sec:exist}, we see that the
value of $p_c(x)\sim m(x)$ also decreases with underdoping. Based on
these trends, we can expect that with underdoping, and hence with
decreasing $T_c$: (i) the resonance energy $\omega_{\rm res}$
decreases and merges with the decreasing spin gap; and (ii) the
incommensurability $\delta_{\rm inc}$ decreases and merges with the
nodal points. These are in qualitative agreement with
experiment~\cite{FonBouSid00a,DaiMooHun01a,ChrMcMRon04b,PaiUlrFau05a}.
We note, in contrast, that if $k_F$ remains fixed while $T_c$
changes in a set of experiments, Eq.~(\ref{eq:dinc}) predicts that
the incommensurability $\delta_{\rm inc}$ will \emph{increase}
linearly with decreasing $T_c$. Underdoping, however, increases
$k_F$, hence the available area for the incommensurate structure
shrinks.

The resonance and the incommensurate features go away at the
transition,  where $m=T_c=0.$ As $T_c\to0$ the resonance merges in
with the continuum response. In our theory, we interpret the absence
of a resonance in La-based cuprates with an anomalously small $T_c$
compared to other cuprate families, as the difficulty to discern
such a merging feature. If this is indeed what is happening, the
same behavior should be seen in other families as $T_c\to 0.$ Of
course, it is possible that the absence of the resonance in
La-cuprates is due to the different physics, that would give rise to
stripes, for example.

To conclude our discussion, we briefly compare our results to existing theoretical literature. The spin response in cuprates has been studied extensively~\cite{Lu92a,BulSca93a,OnuRos95a,BriLee98a,BriLee99a,KaoSiLev00a,ChuJanTch01a,Nor01a,TchNorChu01a,BriLee02a,SchBilMud04a,CheWen05a,PreSeg06a,YamMet06a,Yam07a}. The most common approach to the problem is the random-phase approximation (RPA) derived or postulated in the context of a spin-fermion model such as the (extended) \mbox{$t$-$J$} model. With a $d$-wave gap and by appropriately choosing the hopping amplitudes (fermiology) the resonance arises as the pole of the RPA response at a step (Van Hove) singularity of the bare spin response. The dispersion of the resonance is the result of the dynamical nesting of the energy spectrum~\cite{BriLee02a}. It has also been argued that the $d$-wave gap as opposed to the fermiology is the root of the resonance structure~\cite{KaoSiLev00a,TchNorChu01a}. This resonance has been interpreted as a spin collective mode or exciton. Various phenomenological parameters such as higher harmonics of the superconducting gap~\cite{SchBilMud04a}, interlayer couplings~\cite{CheWen05a,YamMet06a,PreSeg06a}, and Fermi surface geometry~\cite{Nor01a} have been modeled to fit the experimental results.

Clearly, our work shares with the above studies the explanation of the resonance as a spin-collective mode. We calculate the spin response in the ladder approximation which is also the basis of the RPA response above. However, the mechanism that gives rise to the exciton differs in our approach. In fact there is no simple RPA form for our response, and the bare response $\Im\chi_0(\vex p,\omega)\sim\sqrt{\omega^2-\vex p^2}\Theta(\omega^2-\vex p^2)$ does not contain any step singularity~\cite{HerLee03a}. The exciton we find is due to the strong coupling between the spinons and vortex defects of the superconducting order parameter through the Berry gauge field. Perhaps more importantly, the dispersion found here (FIG.~\ref{fig:disp}) persists, at least in our simplified analysis, to very low energies. The dispersion in the RPA approach has a similar shape to ours but practically merges into the continuum at a finite frequency below the commensurate $\omega_{\mathrm{res}}$. In most RPA studies the form of the spectrum away from the nodes (e.g. the curvature of the Fermi surface) is an important part of the explanation. We have, in contrast, only kept the linear part of the spectrum, which is justified within our low-energy theory, and assumed the mass of the Berry gauge field is small enough for the high-energy tail of the gauge-field propagator to be important. Finally, the RPA studies are phenomenological and rely on numerical calculations. They reproduce rather well many aspects of the experiments. Our work is, on the other hand, an analytical study within the effective QED$_3$ theory of underdoped cuprates. To be able to proceed analytically we have relied on various approximations. Although we have argued they are reasonable, these approximations might be too restrictive to contain all the physics revealed in neutron scattering measurements.

\section{Conclusion}\label{sec:conc}

We have studied the problem of particle-hole bound states of spinons (excitons) in the QED$_3$ effective theory of a phase-fluctuating $d$-wave superconductor. This theory contains a massive gauge field with an exact conformal propagator at high-energies. We employed a ladder approximation and derived an approximate Schr\"odinger-like equation for the bound states. We discussed the conditions for the existence of excitons and concluded that they would exist in the strong-coupling regime. We deduced the dispersion of the excitons and compared our results with neutron scattering measurements in cuprates.

This work complements the earlier work of Herbut and Lee~\cite{HerLee03a}, who discuss the continuum spin response in QED$_3$ in a low-energy approximation. It was found there that dispersing incommensurate and commensurate peaks similar to the resonance structure exist in the continuum. The qualitative behaviour of this continuum spectrum remains the same when higher-energy effects are included, but the numerical values of spin gap are reduced and are in fact close to the experimentally-observed range of resonance energies~\cite{Ser06a}. Thus, the continuum spectrum must be accounted for in extracting information on the exciton spectrum from experiments with a finite energy and momentum resolution.

Our vertex function is by definition gauge-dependent. In this work we used a non-local gauge-fixing term, $\xi D(p)\hat p_\mu\hat p_\nu$, in the gauge-field propagator with a momentum-independent parameter $\xi$. However, this is not enough to ensure the gauge invariance of the response for $D(p)$ given by Eq.~(\ref{eq:Dp}): enforcing the Ward-Takahashi identity for the bare spinon propagator yields~\cite{GusHamRee01a} a momentum-dependent parameter $2/3<\xi(p)<1$, with the lower and higher bounds corresponding to the ultraviolet and infrared limits, respectively. A renormalization-group analysis on the original dual theory of Ref.~\onlinecite{Her02b}, from which one can derive the action in Eq.~(\ref{eq:Action}), has been performed in Ref.~\onlinecite{LeeHer03a}. It was found that the charge $g$ cannot be renormalized due to the non-analyticity of the spinon polarization $\sim |p|$. So, within our ladder approximation and for the bare value $g=1$, it is likely that the gauge-invariant response is not in the strong-coupling regime. Further study of the gauge-invariant response seems necessary.

More comprehensive numerical studies of this problem will be very useful. They can be applied, at various stages, to the original vertex equations, or to the Schr\"odinger equation for the bound states. Finally, it is interesting to see whether other spin-collective modes, say triplets, can be formed in QED$_3$. 

\section*{Acknowledgment}
This research was supported by the NSERC of Canada and the CIAR.
 B.H.S. would like to thank M. Case, M. Franz, A. Furusaki, K. Kaveh, and N. Nagaosa for valuable discussions. I.F.H. acknowledges an early collaboration with D. Lee on this project.

\appendix

\section{Derivation of Eq.~(\ref{eq:apprSchr})}\label{app:details}
We decompose the generalized response into symmetric and anti-symmetric components as $\Phi'= \id \phi' + i{\bm{\sigma}}\cdot\phi'_A.$ Then, keeping in mind that $L=r\times p$, from Eq.~(\ref{eq:Up'Schr}) we find
\begin{widetext}
\begin{eqnarray}
&&\left[\frac{p^2}4 - \partial\cdot\partial - \lambda\cosh^2|L/2|\:D(r) \right]\phi'(r,p) + i\frac\lambda2\sinh|L|\:D(r)\:\hat L\cdot \phi'_A(r,p) = \delta^{(3)}(r), \\
&&\left[\frac{p^2}4 - \partial\cdot\partial + \lambda\sinh^2|L/2|D(r)\hat L\hat L\right]\phi'_A(r,p) + i\frac\lambda2\sinh|L|\: D(r)\: \hat L\: \phi'(r,p) = 0.
\end{eqnarray}
\end{widetext}
This is a set of coupled Schr\"odinger equations. However, we notice that the second equation contains a \emph{repulsive} potential and consequently has presumably no negative-energy bound-states. Thus, for $|p|>0$, its Green's function drops exponentially with increasing distance. In order to account for its effects on the bound states of the first equation, we may then solve the second equation for finite $p$ simply as
$$
\phi'_A(r,p) \approx -i\frac{2\lambda}{p^2}\sinh|L|\: D(r)\: \hat L\: \phi'(r,p).
$$
By plugging this expression into the first equation we have $[(p/2)^2 - \partial\cdot\partial - V_{\rm eff}(r) ]\phi'(r,p) = \delta^{(3)}(r)$, where
$$
V_{\rm eff}(r) = \lambda\left[1-\frac{4\lambda}{p^2}\sinh^2|L/2|\:D(r)\right]\cosh^2|L/2|\:D(r).
$$

For $0<|p|<2m$, due to the exponential fall-off of $D(r)$ (see Appendix~\ref{app:rsD}), we have
$$
\frac{4}{p^2}\sinh^2|L/2|\: D^2(r) \approx \frac{|r\times p|^2}{16r^2 p^2}D(r).
$$
Thus, we obtain Eq.~(\ref{eq:apprSchr}). For $|p|>2m$ we still find the same behaviour as above at short distances. But, for large distance, the $\sim\lambda^2$ term overwhelms the other term and we have
$$
V_{\rm eff}(r\gg \frac1{|p|-2m})\sim -\frac{\lambda^2}{p^2}\sinh^2|L|\:D^2(r),
$$
as claimed in Eq.~(\ref{eq:apprSchrL})

\section{Real-space gauge-field propagator}\label{app:rsD}
Here we derive the real-space interaction $D(r)$. Since there is a gap in the low energy limit, we expect an exponential decay in the long-distance behaviour of $D(r)$. Also, the $1/|p|$ tail in the high-energy limit should translate to a $1/r^2$ singularity at short distances. We start by noting that
\begin{eqnarray}
D(r) &=& \frac{m}{4\pi r}\int_0^{\Lambda/2m} F(1/2z)\sin(2m|r|z)\d z \\
       &\equiv& \widetilde{D}(r) - \frac m{8|r|},
\end{eqnarray}
where $\Lambda$ is an ultraviolet cutoff and
\begin{equation}
\widetilde D(r) = \frac{m}{8\pi|r|}\int_{-\Lambda/2m}^{+\Lambda/2m} \frac{1+z^2}{z^2}\tan^{-1}z\sin(2m|r|z)\d z.
\end{equation}

We compute $\widetilde D(r)$ using contour integration. To this end, we need to choose two branch-cuts to define $\tan^{-1}z$. We take them to be $[+i,+i\infty)$ and $[-i,-i\infty)$ such that $z-i = r_1e^{i\theta_1}$ with $-\frac{3\pi}2<\theta_1<\frac\pi2$ and $z+i = r_2e^{i\theta_2}$ with $-\frac{\pi}2<\theta_2<\frac{3\pi}2$. Then,
$$
\tan^{-1}z = -\frac{i\pi}2 +\log\frac{r_2}{r_1} + i(\theta_2-\theta_1).
$$
We take a contour $C$ that includes the real axis and closes on itself in the upper-half plane, except that it avoids the upper branch-cut by tracing the path $C_1=-\digamma(0^+)\cup\digamma(0^-)$ where $\digamma(\varepsilon)=[\varepsilon+i,\varepsilon+i\infty)$. So, we can now write
\begin{eqnarray}
\widetilde D(r) &=& \frac m{8\pi r}\Im\left[\left(\oint_C - \int_{C_1} \right)\frac{1+z^2}{z^2}\tan^{-1}z\: e^{2im|r|z}\d z\right] \nonumber \\
&=& \frac m{8 r}\left[ 1+\int_1^\infty \frac{u^2-1}{u^2}e^{-2m|r|u} \d u \right].
\end{eqnarray}

So, we find
\begin{eqnarray}
D(r) &=& \frac m{8r}\int_1^\infty\frac{u^2-1}{u^2}e^{-2m|r|u}\d u \label{eq:Drint} \\
&=& \frac{e^{-2m|r|}}{16r^2}-\frac{m}{8|r|}E_2(2m|r|). \label{eq:Dr}
\end{eqnarray}
Note that $D(r)>0$. Here, $E_n(x)=\int_1^{\infty}e^{-ux}\d u/u^n$ is the exponential integral.  For $n=2$ it has the following asymptotic behaviour
$$
E_2(x) = \left\{\begin{array}{lr} e^{-x}\left( \frac1x-\frac2{x^2}+\cdots \right), & x\to\infty, \\ 1+(\gamma_E-1+\log x) x + \cdots, & x\to 0, \end{array} \right.
$$
where $\gamma_E$ is the Euler-Mascheroni constant. It leads in turn to the following asymptotic behaviour for $D(r)$:
\begin{eqnarray}
D(r\to\infty) &=& \frac{e^{-2m|r|}}{64m|r|^3}+O(|r|^{-4}e^{-2m|r|}), \\
D(r\to 0) &=& \frac1{16r^2}-\frac m{4|r|}+O(m^2\log(m|r|)).
\end{eqnarray}


\end{document}